# Large Scale (~25 m$^2$) metal diffraction grating of submicron period as possible optoelectronic detector for short scalar gravitational waves

Valery A. Zhukov*

St. Petersburg Institute of Information Science and Automation, Russian Academy of Sciences, St. Petersburg, Russia

## ABSTRACT

A method of detecting of short scalar gravitational waves with a wavelength of λ ~ 0.5 μm is proposed, in contrast to LIGO Project, aimed at detecting of long quadrupole gravitational waves (λ ~ 43 ÷ 10000 km). The conduction electrons in a metal are proposed to use as gravitational receiving antennas (pendulums) instead of massive mirrors in LIGO. It is shown that using a Large Scale metal diffraction grating with area of 25 m$^2$ you can convert the mechanical vibrations of the conduction electrons of metal into a plane electromagnetic wave propagating along the normal to the grating. It is shown that when the amplitude of the scalar gravitational wave in a source (in quasar at the center of our galaxy) is greater than $A_{g0} \approx 5 \cdot 10^{20}$ cm/s$^2$, you can register it with the help of a large optical telescope equipped with the proposed diffraction grating. It is shown that the special theory of relativity allows the amplitude of the scalar gravitational waves in this source by 5 orders of magnitude greater than the above-mentioned minimum value.

## 1. INTRODUCTION

Currently in theoretical astrophysics the prevailing view is that in nature there are only two types of gravitational waves: tensor (quadrupole) and scalar[1]. In recent years, in the frames of LIGO project, attempts were made, yet unsuccessful, to detect long-wavelength (with λ in interval 43÷10000 km) quadrupole gravitational radiation from the millisecond pulsars[2], synchronous with their electromagnetic radiation. It should be noted that the proposed hitherto antennas (detectors) of long wave gravitational radiation are massive pendulums, coupled with sensitive displacement sensors. At the same time there are powerful explosive processes associated with relativistic accretion of matter into a supermassive "black hole" (quasar), located in the center of our galaxy[3-6]. See Figure 1. Processes of accretion in the quasar are accompanied by intense electromagnetic radiation in all observable spectral regions. The power of these processes is superior on the tens of orders the power of processes associated with the rotation of the non-axisymmetric pulsars[5,6]. As it follows from[1], the scalar gravitational waves should also be emitted at such accretion. Moreover, it must be in sync with electromagnetic bursts and in all regions of the spectrum, including the optical frequencies.

## 2. STATEMENT OF THE PROBLEM

In this paper we propose a detector of scalar gravitational waves with a length of λ ~ 0.5 microns. In the last stage of registration of gravitational waves the energy of these waves is converted into the energy of electromagnetic waves of the same frequency. Therefore, the wavelength is selected with the expectation that, as a receiver of these oscillations will be used optical telescope (the tool that is the most sensitive for them). The detector - converter, proposed by us, is a flat metal film with the thickness of the order of λ. We will place this film so that the wave vector of gravitational waves, whose direction is taken as the X-axis of our coordinate system, will lie in the plane of the film. The electrons in the conduction band of the metal are used in this detector as pendulums. By analogy with the Mossbauer Effect we believe that the ion subsystem of metal at temperatures below the Debye temperature will not respond to the impact of short

*valery.zhukov2@gmail.com

gravitational waves. As example, we give Debye temperature for some metals[7] in Table 1. For reasons of symmetry the scalar gravitational wave is a longitudinal wave of vibrations of the gravitational potential gradient, such as a wave of longitudinal oscillations of the particles in an acoustic wave in the gas. Therefore, we believe that the scalar gravitational

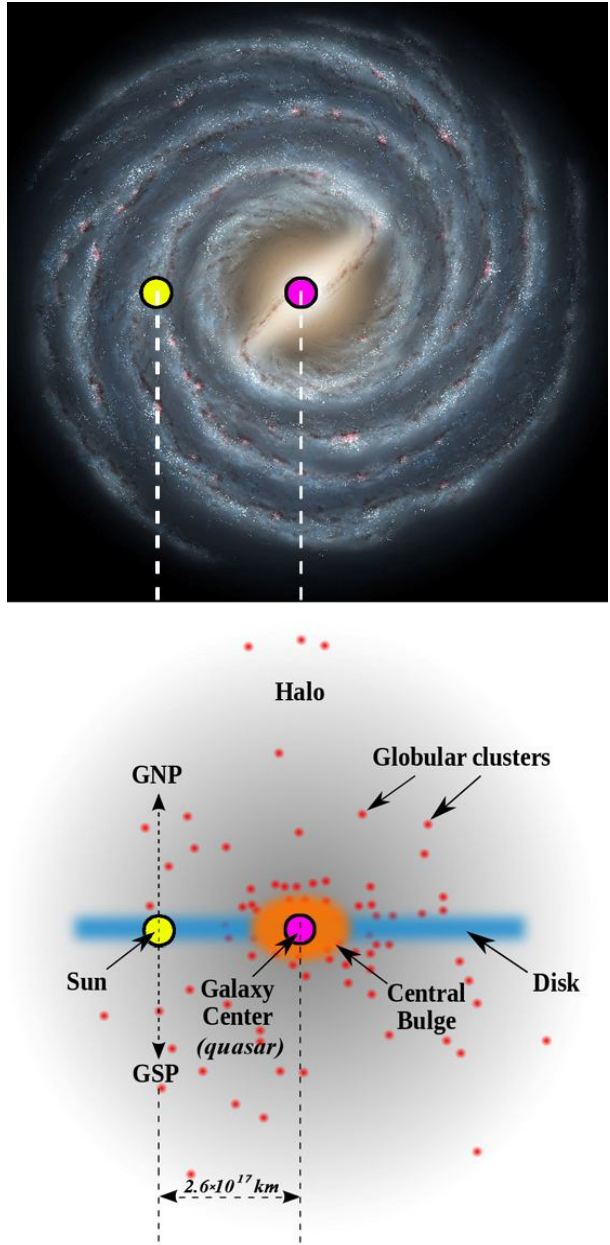

Figure 1. Milky Way Galaxy (front view at the top, side view at the bottom): mutual position of the quasar and the Sun.

Table 1. Debye's Temperature $\Theta_D$ in Kelvin for some metals.

| metal | Al | Cu | α- Fe | Ti | W | Ni |
|---|---|---|---|---|---|---|
| $\Theta_D(K)$ | 429 | 344 | 464 | 420 | 405 | 440 |

wave in the linear approximation (weak field) Galileo metric[8] is mathematically described by the formula

$$A_g(R,t) = (A_{g0}R_g / R)\cos(\omega t - kR),  \tag{1}$$

where $A_g(R,t)$ is the value of the potential gradient of variable gravitational field at a distance $R$ from the source (the supermassive "black hole" in the center of the Galaxy[3-6]) at time t. In our case $R \approx 8.33$ kpc (or converting to kilometers $R \approx 2.6 \cdot 10^{17}$ km), $A_{g0}$ is the peak value of the alternating gradient of the gravitational field in the source at a distance equal to the Schwarzschild radius from the center. This radius, otherwise known as the gravitational radius of the "black hole" $R_g$ is equal $R_g = 2\gamma M / c^2$, where $\gamma = 6.67 \cdot 10^{-11}$ m$^3$·kg$^{-1}$·s$^{-3}$ ($\gamma = 6.67 \cdot 10^{-8}$ cm$^3$·g$^{-1}$·s$^{-3}$) is the gravitational constant. For a "black hole" in the center of the galaxy having a mass $M_c \approx 4.31 \cdot 10^6 M_{Sun}$, where $M_{Sun}$ is the mass of the Sun, $R_g \approx 12 \cdot 10^6$ km. On considered distances from the source we can approximately assume that the spherical gravitational wave is flat[8] and the formula (1) takes the form

$$A_g(x,t) = (A_{g0}R_g / R)\cos(\omega t - kx),  \tag{1'}$$

where x-coordinate is measured along the direction of the wave propagation. In the quantum theory of solids it is assumed that the motion of electrons in the conduction band of the metal is described by a collective wave function composed of individual electron orbitals, the so-called periodic Bloch functions[9].

When considering the process of reflection of weak electromagnetic waves from the surface of a metal mirror, it is believed that additional motion under the influence of the electromagnetic wave is linearly superimposed on the orbital motion of electrons. This additional motion of charges, in turn, stimulates the secondary (mirrored) electromagnetic wave[10]. By analogy with the process of reflection of electromagnetic waves by electrons of the metal we consider the process of interaction with them a gravitational wave. The electrons in the conduction band of the metal, by virtue of the principle of equivalence of inertial and gravitational masses, under the influence of gravitational waves, in addition to their motion on Bloch orbitals will oscillate along the direction of wave propagation with an acceleration equal to the gradient of the potential in the wave: $A_g(x,t)$. Each electron with its nearest metal ion of the crystal lattice can be considered as an elementary harmonic oscillator - Hertz dipole. This is done, for example, when calculating the bremsstrahlung X-radiation[11]. This dipole can be described by a function of the form

$$p(x,t) = q_e \cdot l(x,t) = p_0 \cdot f(x,t),  \tag{2}$$

where $p_0 = q_e \cdot l_0$, $q_e$ is electron charge, $l_0$ is the maximum amplitude of the deviation from the equilibrium position of the electron. (Here and below we follow the presentation[8-11] and will use a system of units CGSE.) From the formulas (1') and (2) and the principle of equivalence of inertial and gravitational masses it follows:

$$\ddot{l}(x,t) = A_g(x,t) = A_{g0}\cos(\omega t - kx) \cdot (R_g / r)  \tag{3}$$

Assuming $l(x,t) = l_0 \cos(\omega t - kx)$ we get $\ddot{l}(t) = -l_0 \omega^2 \cos(\omega t - kx)$. Hence, in turn, it follows

$$l_0 = \left| A_{g0}(R_g / r)/\omega^2 \right|  \tag{4}$$

The distances between the individual Hertz dipoles (conduction electrons in metals) are equal approximately to the lattice constant, thus it is about of 0.5 nm (5 * 10$^{-8}$ cm). In result of acceleration transmitted to electrons from gravitational waves these dipoles will radiate electromagnetic waves. The characteristic frequency of the oscillations in the emitted electromagnetic wave remains the same, and the speeds of propagation of electromagnetic and gravitational waves in vacuum coincide. Thus wave number $k$ for the searched electromagnetic wave is equal to wave number $k$ for the gravitational wave from the source. According to[11], outside of the metal film, at a distance r from it, which satisfies the condition of far zone ($kr \gg 1$), these oscillations are described by the formula:

$$\vec{E} = k^2 [\vec{n} \times [\vec{n} \times \vec{p}(x,t)]] / r = k^2 [\vec{n} \times [\vec{n} \times (q_e \cdot l_0 \cos(\omega t - kx)\vec{e}_x + 0 \cdot \vec{e}_y + 0 \cdot \vec{e}_z)]] / r .  \tag{5}$$

It is easy to see that $\vec{E} = E_x \vec{e}_x + 0 \cdot \vec{e}_y + 0 \cdot \vec{e}_z$, and $E_x = E_0 \cos(\varpi t - kx) = \varphi_0 \cos(\omega t - kx)/r$, where $E_0 = \varphi_0 / r$ and $\varphi_0 = k^2 q_e \cdot l_0$. Simultaneously the fluctuations of the magnetic field will occur[11] that will be described by the formula $\vec{B} = [\vec{n} \times \vec{E}]$.

Consider the picture of the electric field distribution at a fixed time t. For each area of the metal film for which, according to formula (5), the vector $\vec{E}$ is directed along the direction of the gravitational waves will match the area for which the vector $\vec{E}$ is directed in the opposite direction. In space above the metal film the values of the electric field vector, for which the sources are in these areas, when averaged give a zero result.

Consider on metal film two adjacent parallel strips that are perpendicular to the direction to the source of gravitational waves, and separated by a distance of $\lambda/2$. See Figure 2.

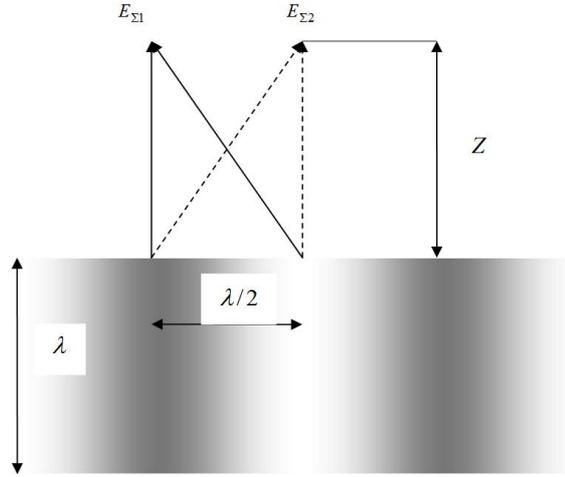

Figure 2. Electric fields $E_{\Sigma 1}$ and $E_{\Sigma 2}$ above two dedicated strips on flat metal film.

At a distance z, deferred from the film perpendicular to it, the total electric field from the two dedicated strips above the first strip is: $E_{\Sigma 1} = \varphi_0 \cos(\omega t - kx)\left(1/z - 1/\sqrt{z^2 + \lambda^2/4}\right) \propto \varphi_0 \cos(\omega t - kx)\lambda^2/8z^3$.

The total electric field above the second stripe is equal to:
$E_{\Sigma 2} = \varphi_0 \cos[\omega t - k\cdot(x+\lambda/2)]\left(1/z - 1/\sqrt{z^2 + \lambda^2/4}\right) \propto -\varphi_0 \cos(\omega t - kx)\lambda^2/8z^3$. I.e. field decreases as $1/z^3$. The electromagnetic field above our metal film in the "far" zone (at $kr \gg 1$) for considered electromagnetic vibrators will be absent.

## 3. METHOD

For arising of this electromagnetic field, you need to make an orderly change on $\pi$ of the phase of the oscillations generated in the metal in the areas in which the electrons oscillate in opposite directions. For this purpose, we will make the surface of our metal film corrugated so that its coordinate along axis Z direction, perpendicular to the film plane will vary in accordance with formula

$$z(x) = \lambda/4\,(1+\cos kx) + 3\lambda/4.  \qquad (6)$$

Here the origin of the Z - coordinate is already in the bottom surface of the metal film. This is reflected in Figure 3.

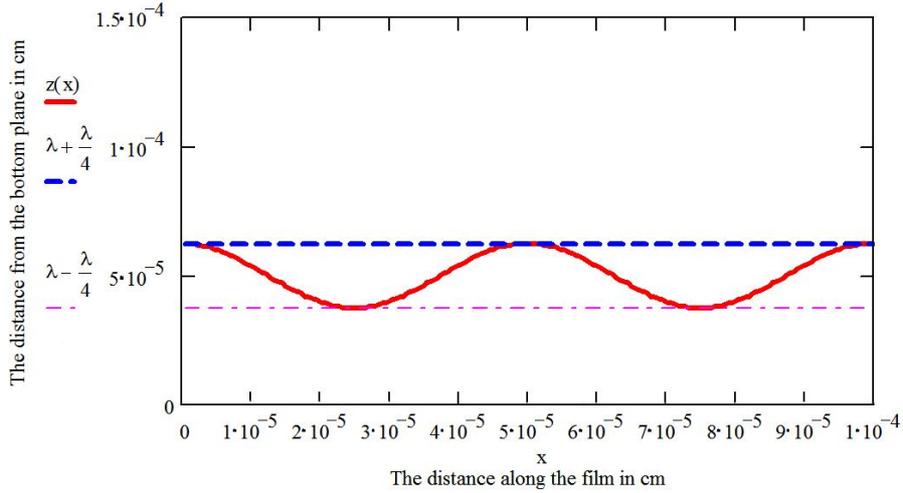

Figure 3. Cosine shape profile of the metal diffraction grating.

Suppose that for the considered gravitational waves the characteristic coherence time exists during which the wave can be considered as stationary. As is well known[12,13], the stationary wave field from sources distributed in the plane XY, at a short distance z from this plane, in the case when diffraction can be neglected is given by the expression

$$E(x, y, z, t) = E(x, y) \cos \omega t \cos kz ,\qquad(7)$$

where the factor $\cos kz$ characterizes the wave propagation in the direction Z, perpendicular to XY plane. In this case,

$$E(x, y, z, t) = E(x, z, t) = E_0 \cos(\omega t - kx) \cos kz ,\qquad(7')$$

and $E_0 = k^2 q_e \cdot l_0 / r$ according to formula (5). Then, in a plane passing through the top of the folds of our corrugated surface which is predetermined by the formula (6), the expression (7') is transformed into:

$$E_0 \cos(\omega t - kx)\cos kz = E_0 \cos \omega t \cos kx \cos(\pi/2 (1+\cos kx)) - E_0 \sin \omega t \sin kx \cos(\pi/2 (1+\cos kx)) .\qquad(8)$$

We introduce the notation: $f_1(x) \equiv \cos kx \cos(\pi/2 (1+\cos kx))$ and $f_2(x) \equiv \sin kx \cos(\pi/2 (1+\cos kx))$. Then the formula (8) can be rewritten as:

$$E(x, z, t) = E_0 \cos \omega t \cdot f_1(x) - E_0 \sin \omega t \cdot f_2(x) .\qquad(8')$$

Figure 4 and Figure 5 show graphs of functions $f_1(x)$ and $f_2(x)$ as well as their simplest approximation functions $1/2 + \cos(2kx)/2$ and $\sin(2kx)/\sqrt{2.3}$, accordingly.

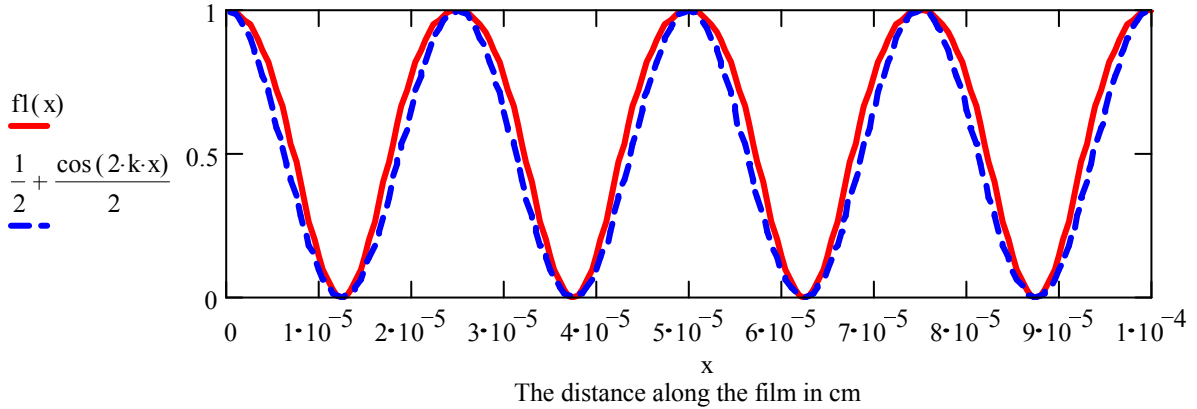

Figure 4. The solid red line is graph of the function $f_1(x)$; for comparison: the blue dashed line in the same Figure is graph of the function $1/2 + \cos(2kx)/2$.

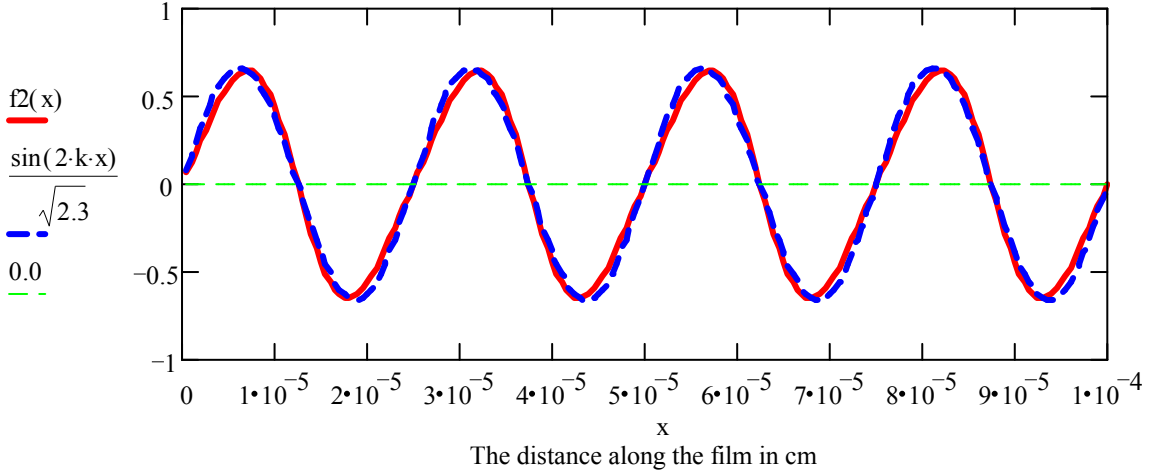

Figure 5. The solid red line is the graph of the function $f_2(x)$; for comparison, dashed blue line in the same Figure is the graph of the function $\sin(2kx)/\sqrt{2.3}$.

Standard deviations of functions $f_1(x)$ and $f_2(x)$ for the relevant comparison functions were calculated:

$$\sqrt{\int_0^\lambda [f1(x)-(1+\cos 2kx)/2]^2\,dx \Big/ \int_0^\lambda [(1+\cos 2kx)/2]^2\,dx} = 0.134 \tag{9}$$

$$\sqrt{\int_0^\lambda [f2(x)-\sin 2kx/\sqrt{2.3}]^2\,dx \Big/ \int_0^\lambda [\sin 2kx/\sqrt{2.3}]^2\,dx} = 0.114 \tag{9'}$$

As can be seen from the graphs in Figure 4 and formula (9), the first term on the right-hand side of (8') is in good agreement with the function $E_0 \cos\omega t(1/2+\cos(2kx)/2)$. This function, in turn, can be divided into two components: $(E_0/2)\cos\omega t$ and $(E_0/2)\cos\omega t\cos(2kx)$. Comparison with formula (7) shows that first component, the function $(E_0/2)\cos\omega t$ will correspond to a plane electromagnetic wave. This wave has a constant within the wave front amplitude of the electric vector, equal to $E_0/2$ and propagates normal to the metal film. Consider now the second component in the approximating function that shown in Figure 4 - function $(E_0/2)\cos 2kx$ depending on the variable x. We get a field $(E_0/2)\cos\omega t\cos 2kx$, taking also into account depending on the time. From the theory of oscillations is known that the sum of two waves propagating in opposite directions, namely $\cos(\omega t+kx)+\cos(\omega t-kx)=2\cos\omega t\cos kx$, is a standing wave. In our case it will correspond to the "pseudo-standing" electromagnetic wave "propagating" parallel to the metal film in a fictitious medium with refractive index $n=2$. Such waves will not be radiated into the environment. The second term in equation (8') that is approximated by function $E_0\sin\omega t\sin 2kx/\sqrt{2.3}$ (see Figure 5), will also match "pseudo standing" wave "propagating" parallel to the film. Thus, only about a third of the energy transmitted by a gravitational wave to electron subsystem of the metal film to be transformed into the energy of the electromagnetic wave, propagating out of the film.

## 4. RESULTS

A plane light wave obtained in our antenna should be directed to the parabolic mirror of modern large optical telescope parallel to the optical axis of its main mirror and then to a photo detector of telescope, see Figure 6. Figure 7 shows the optical (radio) telescope before (upper panel a)) and after (bottom panel b)) of its conversion into a gravitational telescope. At the bottom panel b) the flat metal membrane (film) is shown, that is stretched perpendicular to the optical axis of the telescope. "Finishing touches" of the diffraction grating are shown on this membrane for clarity conditionally (not to scale). In reality these "touches" should be towards the surface of the telescope mirror. Gravitational wave $A_g(x,t)$ (see formula (1')) incident on our "antenna" (diffraction grating) must excite in it a light wave with a power greater than the limit sensitivity of the photo detector of the optical telescope $P_{\min}$.

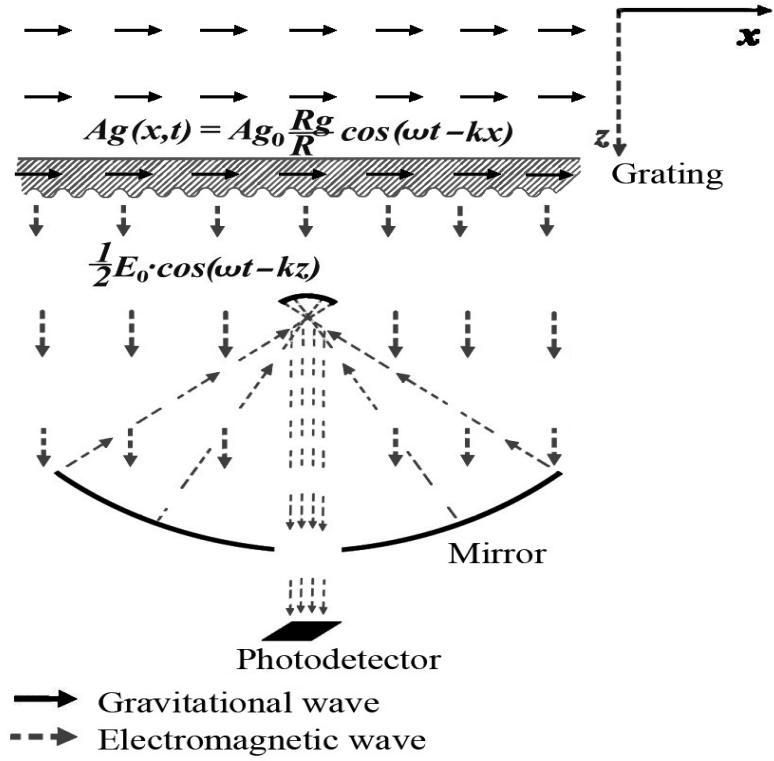

Figure 6. Scheme of gravitational telescope.

We now define the lower limit for the amplitude of the scalar gravitational wave in source $A_{g0}$, which may be registered by means of the obtained gravitational telescope. We use the formula out of[11], which gives the time-averaged power of the electromagnetic energy emitted by the elementary Hertz dipole in all directions

$$\overline{P}_t = \omega^4 p_0^2 /(3c^3) = (2\pi/\lambda)^4 (p_0^2 c/3) . \qquad (10)$$

In our case only part of this power will be radiated outside the antenna. Firstly, it will be part attributable to half of the total solid angle $4\pi$ radian. Secondly, it will be one third of that half because of the incomplete conversion of the remaining part of the radiation into the plane wave with the wave surface that is parallel to the plane of the antenna. As a result, in all subsequent formulas for the radiated electromagnetic energy the factor 1/6 will be needed to enter. Since our gravitational antenna emits planar light wave, the area of the antenna in an optimal case must be equal to the area S of the used telescope mirror. Given that the depth of the attenuation of electromagnetic waves in the metal ~ λ, the working volume occupied in our antenna by conduction electrons, emitting under the influence of gravitational waves having a wavelength λ, is equal to $V = S\lambda$. The total number N of electrons involved in the process, is equal $N = n_e S\lambda$, where $n_e$ is number of electrons in the conduction band per unit volume. The total radiated power must satisfy the condition

$$(\overline{P}_t /6) \cdot N = (2\pi/\lambda)^4 p_0^2 c/18 \cdot n_e S\lambda \geq P_{\min} , \qquad (11)$$

where $P_{min}$ is the limit sensitivity of the optical telescope. On the basis of this formula using the formula (5), first we will find the lower limit of the allowable value of $A_{g0}$ - the amplitude of the scalar gravitational wave in a source allowing to register it using proposed detector. It is expressed by the formula

$$A_{g0} = (R/eR_g)\sqrt{18c^3 P_{\min}/(\lambda n_e S)} . \qquad (12)$$

In the available literature there are no direct data on the limit sensitivity of the optical telescope $P_{min}$. But we can estimate it by indirect data about extremely distant stars detected by modern large telescopes. To date[14], it is a "red giant" that is separated from the solar system by a distance of $R_{rg} = 400000$ light years. Suppose, according to[15], that the luminosity of such star is equal $\sim 10^2 L_\Theta$, where $L_\Theta = 2.8 \cdot 10^{26}$ Watt is luminosity of the Sun. Then the incident radiation from the "red giant" on the mirror of the telescope with diameter of $D_t = 10$ m has power equal to

$$P_{\min} = 10^2 L_\Theta D_t^2 /(16 R_{rg}^2). \qquad (13)$$

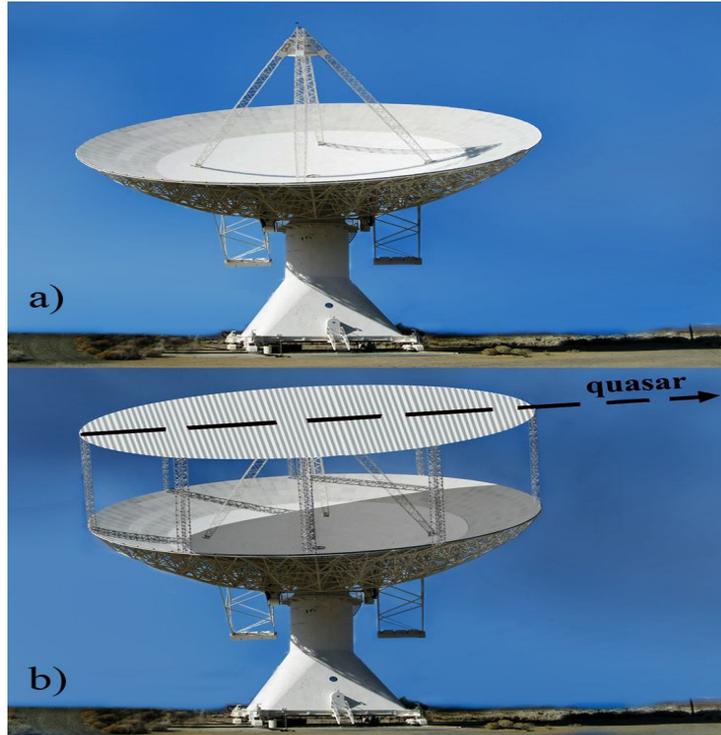

Figure 7. a) The conventional telescope with parabolic mirror. b) The gravitational telescope with the diffraction grating above the parabolic mirror.

Substituting in this formula the values of all variables, we obtain for the limit sensitivity of the optical telescope the value: $P_{\min} \approx 10^{-14}$ W or in the system CGS: $P_{\min} \approx 10^{-7}$ ergs/sec. For reliable detection of gravitational wave the power of a plane light wave emitted by gravitational antenna on the mirror of the telescope should be greater than this value. Substituting this value $P_{\min}$ into (12), we obtain that the value of the amplitude of the gravitational wave in the source is equal to $A_{g0} \approx 5 \cdot 10^{20}$ cm/s². We now estimate the maximum value of the amplitude at the source, the allowable accordingly special relativity. To do this, we check does not exceed the peak value of the velocity of the body, which is the source of this wave, the speed of light. We assume that the gravitational wave is the result of a quasi-periodic process with a characteristic angular frequency $\omega = c \cdot 2\pi / \lambda$. Then, amplitude value of the velocity is expressed by the formula $v_{ampl} = A_{g0} / \omega$ and in this case it is equal to $v_{ampl} \approx 1.5 \cdot 10^5$ cm/sec at the speed of light in the CGS equal to $c \approx 3 \cdot 10^{10}$ cm/sec. Thus we have the limit value of the amplitude of the gravitational wave in the source equal to $A_{g0} \approx 10^{26}$ cm/s² and margin of safety allowed by the theory of special relativity in 5 additional decimal orders.

## 5. CONCLUSION

Suppose now that in the vicinity of the solar system, not exceeding 1 light year, there is a "black hole" of stellar mass. And let in such "black hole" the processes of accretion occur. Let us now state the question whether it is possible to register the gravitational waves emitted by such "black hole" using the proposed antennas? Let the mass of a black hole is approximately equal to 2.5·$M_{Sun}$, then its Schwarzschild radius $Rg$ is equal to ~ 5 km. Consider the case of a gravitational antenna combined with an optical telescope. For this case, the minimum amplitude of the gravitational wave in a source, that is sufficient to detect the signal, is equal: $A_{g_0} \approx 4 \cdot 10^{22}$ cm/s$^2$, and the maximum value, based on the special theory of relativity, is equal: $A_{g_0 \lim} \approx 0.6 \cdot 10^{26}$ cm/s$^2$. Thus we have the margin of safety about of 3 decimal orders of magnitude. Now imagine a situation when into the Solar system from the direction opposite to that where we see a star Proxima Centauri, at a distance of ~ 1 light year, with a speed of 500 km/sec that is possible for former kernel of supernovae, a wandering black hole is moving. In this case, there is a chance to detect it using the suggested tools and it remains about 600 years for humanity for thinking about what to do with this information.

The following method can be proposed for manufacturing of the above-described diffraction grating with area of 25 m$^2$. Take a square polished silicon wafer with size 200x200 mm$^2$. Create the desired metal profile on it by photolithography and plasma-ion etching. And then to put together the 625 such plates on flat ground.

## ACKNOWLEDGMENTS

This work was supported by the Program of Fundamental Investigations of the Presidium of the Russian Academy of Sciences no. 24 "Foundations of Basic Research of Nanotechnologies and Nanomaterials", Section "Physics of Nanostructures and Nanoelectronics" and by the Program of Basic Scientific Research of the Department of Nanotechnologies and Information Technologies of the Russian Academy of Sciences "Elemental Base of Microelectronics, Nanoelectronics, and Quantum Computers; Materials for Micro- and Nanoelectronics, Microsystem Technique, Solid_State Electronics".

## REFERENCES

[1] Hajime Sotani, "Scalar gravitational waves from relativistic stars in scalar-tensor gravity", arXiv:1402.5699v1
[2] http://www.ligo.org/science/outreach.php
[3] Majaess, D., "Concerning the Distance to the Center of the Milky Way and Its Structure" Acta A. 60 (2010).
[4] Melia Fulvio, [The Black Hole in the Center of Our Galaxy], Princeton U Press, 2003.
[5] Eckart, A., Schödel, R., Straubmeier, C., [The Black Hole at the Center of the Milky Way], London: Imperial College Press, 2005.
[6] Melia Fulvio, [The Galactic Supermassive Black Hole], Princeton U Press, 2007.
[7] Launay, J. De., [Solid State Physics], Vol., 2 ed. Seitz, F. and Turnbull, D., (Academic Press, New York, 1956).
[8] Landau, L.D., Lifshitz, E.M., [The Classical Theory of Fields] (Volume 2 of A Course of Theoretical Physics), Pergamon Press 1971.
[9] Kittel, C., [Quantum Theory of Solids], John Wiley & Suns, Inc. New York-London, 1963.
[10] Born, Max and Wolf, Emil, [Principles of Optics], Pergamon Press, Oxford-London-Edinburg-New York-Paris, 1964.
[11] Terletskii, Ya. P., Rybakov, Yu. P., [Elektrodynamika (Electrodynamics)], «Vysshaya Shkola" Publishing, Moscow, (1980) (in Russian).
[12] Haus, H. A., [Waves and Fields in Optoelectronics], Prentice-Hall, Inc. Englewood Cliffs, New Jersey, 1987.
[13] Cowley, J. M., [Diffraction Physics], North-Holland Publishing Company, Amsterdam-Oxford , 1975.
[14] Bochanski, J. J., Willman, B., Caldwell, N., Sanderson, R., West, A. A., Strader, J., Brown, W., "The Most Distant Stars in the Milky Way", The Astrophysical Journal Letters, Volume 790, Issue 1, article id. L5, pp. 6, 2014.
[15] Zeilik, Michael A., Gregory, Stephan A., [Introductory in Astronomy & Astrophysics] (4th ed.), Saunders College Publishing, pp. 321–322. ISBN 0-03-006228-4, 1998.